# *Engineering Nonreciprocal Responses in Travelling-Wave Spacetime Crystals via Clausius-Mossotti Homogenization*

**Filipa R. Prudêncio[1,2*], and Mário G. Silveirinha[1]**

[1]University of Lisbon – Instituto Superior Técnico and Instituto de Telecomunicações, Avenida Rovisco Pais 1, 1049-001 Lisbon, Portugal

[2]Instituto Universitário de Lisboa (ISCTE-IUL), Avenida das Forças Armadas 376, 1600-077 Lisbon, Portugal

## Abstract

Here, we investigate the effective response of three-dimensional spacetime crystals formed by spherical scatterers under a travelling-wave modulation. We develop an analytical formalism to homogenize the spacetime crystals that extends the renowned Clausius-Mossotti formula to time-varying platforms. Our formalism shows that travelling-wave spacetime crystals can be used to engineer a wide range of classes of nonreciprocal bianisotropic couplings in the long wavelength limit. In particular, our theory reveals the possibility of realizing a purely isotropic Tellegen response in crystals formed by interlaced sub-lattices of scatterers subjected to different modulation velocities. Furthermore, we introduce a class of generalized Minkowskian crystals that displays invariance under arbitrary Lorentz boosts aligned with a fixed spatial direction. We prove that such systems are formed by pseudo-uniaxial materials with the principal axis aligned parallel to the modulation velocity. The electromagnetic response of such generalized Minkowskian crystals is indistinguishable from that of moving photonic crystals.

---

[*] Corresponding author: filipa.prudencio@lx.it.pt

# I. Introduction

In recent years, the physics and applications of time-varying systems have garnered significant interest [1-20]. Time modulations offer powerful knobs to control the material response in unique ways, and can be useful to design unidirectional guides and isolators [4-16], optical amplifiers, oscillators and for frequency-multiplication [2-19].

In "travelling-wave" spacetime modulated systems the variations in space and time of a system are locked in such a way that the material response is only a function of $\mathbf{r} - \mathbf{v}t$ where $\mathbf{r} = (x, y, z)$ is the observation point, $t$ is time and $\mathbf{v}$ is the modulation speed [20-22]. Several previous works have analyzed the effective response of 1D type (laminate) spacetime crystals in the long wavelength limit [21-26]. Most notably, the laminate systems may exhibit an effective nonreciprocal magneto-electric coupling [20, 21, 22, 27]. When the spacetime crystal is formed by isotropic materials the effective magneto-electric tensor is anti-symmetric, consistent with a moving-medium type response [20, 21, 22]. Magneto-electric tensors with more general structures can be engineered by considering anisotropic inclusions [22, 27]. In particular, it has been shown that by controlling the main axes of the permittivity and permeability tensors it is feasible to design systems with an effective anisotropic Tellegen response [22].

Here, we go one step beyond these previous studies, and characterize the effective response of fully 3D travelling wave spacetime crystals. For simplicity, we restrict our attention to the case where the spacetime crystal is formed by a cubical array of spherical-type inclusions subjected to a travelling wave modulation. Using a quasi-static formalism, we derive the exact polarizability of a generic spherical particle in the Lorentz co-moving frame coordinates. Then, by combining this result with well-known (Lorentz-Lorenz) local field formulas, we derive a generalized Clausius-Mossotti formula for the effective response of the spacetime crystal. The classical Clausius-Mossotti approach relates the effective permittivity of a composite with the electric polarizability of its constituent elements [28, 29, 30]. It is widely used in the literature for the modeling of metamaterials and other composites [29-35].

Using the developed theory, we investigate the most general class of electromagnetic responses that can be engineered with travelling-wave spacetime crystals, supposing that the inclusions in the laboratory frame are described by arbitrary symmetric permittivity and permeability tensors. Furthermore, we extend our homogenization theory to time-varying crystals formed by interlaced lattices of scatterers subjected to different modulation velocities. We prove that by combining three different sub-lattices formed by elements modulated along directions parallel to the three coordinate axes it is possible to engineer an ideal isotropic magneto-electric Tellegen (axion) response [36, 37]. The axion coupling has received great attention in recent years due to its connections to dark matter and to the topological magneto-electric effect [38-42]. Practically, it may lead to the development of nonreciprocal devices that operate without an external magnetic bias [43-45]. Axion-type responses can emerge spontaneously in nature in some antiferromagnetic materials [46] and in electronic topological insulators [40-42]. Some recent proposals to realize the axion coupling are based on arrays of gyrotropic particles [47-51], and spacetime crystals [22]. The Tellegen-coupling could be useful for engineering nontrivial topological photonic phases and topologically protected surface states [52-54].

It is well-known that the response of a spacetime crystal can be rendered time-invariant through a suitable (Galilean or Lorentz) coordinate transformation [12, 21, 22]. Generally, the material matrix structure in the transformed coordinate system (co-moving frame) differs from that in the original coordinate system (laboratory frame). Specifically, the material response in the co-moving frame is typically characterized by magneto-electric bianisotropic coupling, complicating the study of wave propagation. Recently, we introduced a particular subclass of spacetime crystals whose material response remains unchanged under any Lorentz boost [55]. These systems, termed Minkowskian spacetime crystals, are composed of isotropic isorefractive materials. It was shown in Ref. [55] that the dispersion of Minkowskian spacetime crystals can be characterized straightforwardly using a relativistic Doppler transformation.

In this work, we extend the analysis of Ref. [55] to a broader class of spacetime crystals. Instead of requiring the invariance of the material response for arbitrary Lorentz boosts, we only require response invariance for Lorentz boosts aligned with the modulation speed direction. This relaxed condition ensures that the wave propagation in the laboratory frame can be derived from the wave propagation of a conventional photonic crystal (without magneto-electric coupling) using a relativistic Doppler transformation. We show that a general class of dielectric reciprocal media invariant under Lorentz boosts along a specific spatial direction is comprised of "pseudo-uniaxial" media. In these systems, the tensor $\overline{\overline{\mu}}^{-1} \cdot \overline{\overline{\varepsilon}}$ exhibits a uniaxial structure (with two degenerate eigenvalues), with the optical axis aligned along the Lorentz boost direction.

The article is organized as follows. In section II, we extend the Clausius-Mossotti homogenization to 3D spacetime crystals with a travelling wave modulation. In particular, we present a solution to implement an isotropic Tellegen-type response with 3D spacetime crystals. In Section III, we introduce a general class of dielectric materials whose electromagnetic response is insensitive to Lorentz boosts along a fixed spatial direction. We numerically characterize the exact band structure of 3D Minkowskian spacetime crystals and demonstrate that our homogenization theory accurately describes the wave dispersion in the long wavelength limit. Finally, section IV summarizes the main findings.

## II. Homogenization of spacetime crystals formed by spherical particles

### A. *Spacetime crystal of spherical particles*

We consider the problem of wave propagation in a 3D spacetime crystal formed by a periodic array of spheroid particles embedded in a uniform dielectric [Fig. 1]. In the laboratory frame, the constitutive relations are of the type:

$$\begin{pmatrix} \mathbf{D}(x,y,z,t) \\ \mathbf{B}(x,y,z,t) \end{pmatrix} = \mathbf{M} \cdot \begin{pmatrix} \mathbf{E}(x,y,z,t) \\ \mathbf{H}(x,y,z,t) \end{pmatrix}. \qquad (1)$$

Here, $\mathbf{M}$ is the material matrix defined by

$$\mathbf{M}(x,y,z,t) = \begin{pmatrix} \varepsilon_0 \bar{\varepsilon}(x,y,z-vt) & 0 \\ 0 & \mu_0 \bar{\mu}(x,y,z-vt) \end{pmatrix}. \qquad (2)$$

This matrix comprises the permittivity and permeability tensors, which exhibit variations in both space and time. These variations are synchronized by a constant modulation speed $v$ oriented along the $z$-axis. Consequently, the material response is subject to a traveling wave modulation.

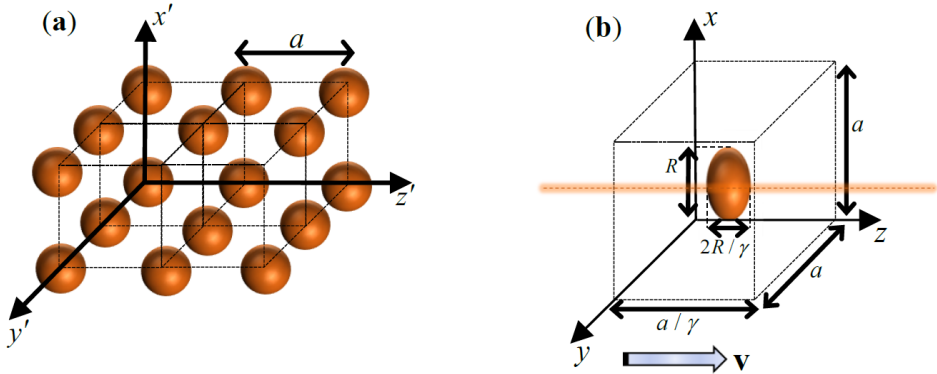

**Fig. 1** Geometry of a spacetime crystal formed by spheroid type inclusions. **(a)** Geometry of the crystal in the Lorentz co-moving frame coordinates. The inclusions are spherical with radius $R$ and are arranged in a cubic array with period $a$. **(b)** Unit cell as seen in the laboratory frame coordinates. Due to the Lorentz-Fitzgerald contraction, the inclusions are slightly ellipsoidal and the unit cell is slightly contracted along the $z$-direction. The horizontal line shaded in orange represents the path traversed by the inclusion.

To simplify the analysis, it is convenient to switch to a new coordinate system where the material parameters become time invariant. It is advantageous to adopt a coordinate transformation that ensures the host region's response is unaffected by the transformation. This can be achieved with a suitable generalized Lorentz transformation [55]. For simplicity, we assume here that the host region is air, which is invariant under the standard Lorentz transformation.

In the new frame (co-moving frame), the spheroid inclusions have a bianisotropic response described by a time-independent material matrix of the form [22]:

$$\mathbf{M}' = \begin{pmatrix} \varepsilon_0 \overline{\varepsilon}' & \frac{1}{c}\overline{\xi}' \\ \frac{1}{c}\overline{\zeta}' & \mu_0 \overline{\mu}' \end{pmatrix}. \tag{3}$$

In the above, $\overline{\varepsilon}', \overline{\mu}'$ are the transformed permittivity and permeability tensors in the co-moving frame and $\overline{\xi}' = \left[\overline{\zeta}'\right]^{\mathrm{T}}$ are the magneto electric tensors. The superscript T represents the matrix transpose operation. As already mentioned, the response of the host region is unaffected by the Lorentz transformation. The primed material matrix is derived from the material matrix of the inclusions in the laboratory frame $\mathbf{M}$ [22, Appendix A]:

$$\mathbf{M}' = \left[\frac{1}{c^2}\mathbf{V} + \mathbf{A}\cdot\mathbf{M}\right]\cdot\left[\mathbf{A} + \mathbf{V}\cdot\mathbf{M}\right]^{-1}. \tag{4}$$

Here, $\mathbf{A}, \mathbf{V}$ are 6×6 matrices written in terms of the modulation velocity ($\mathbf{v} = v\hat{\mathbf{z}}$) and of the Lorentz factor $\gamma = \left(1 - v^2/c^2\right)^{-1/2}$:

$$\mathbf{A} = \begin{pmatrix} \gamma\mathbf{1}_\perp + \hat{\mathbf{u}}_\parallel \otimes \hat{\mathbf{u}}_\parallel & 0 \\ 0 & \gamma\mathbf{1}_\perp + \hat{\mathbf{u}}_\parallel \otimes \hat{\mathbf{u}}_\parallel \end{pmatrix}, \tag{5a}$$

$$\mathbf{V} = \begin{pmatrix} 0 & \gamma\mathbf{v}\times\mathbf{1}_{3\times 3} \\ -\gamma\mathbf{v}\times\mathbf{1}_{3\times 3} & 0 \end{pmatrix}, \tag{5b}$$

where $\mathbf{1}_\perp = \mathbf{1}_{3\times3} - \hat{\mathbf{u}}_\parallel \otimes \hat{\mathbf{u}}_\parallel$, $\mathbf{1}_{3\times3}$ is the identity matrix, and $\otimes$ represents the tensor product of two vectors. The symbols $\parallel$ and $\perp$ refer to the field components parallel and perpendicular to the velocity, so that $\hat{\mathbf{u}}_\parallel \equiv \hat{\mathbf{z}}$.

Importantly, because of the Lorentz-Fitzgerald length contraction [28, 55] the geometry of the photonic crystal in the laboratory and co-moving frames differs by a scaling factor determined by the Lorentz factor $\gamma$. Specifically, the dimension of the spheroid inclusions along the $z'$ direction is enlarged by $\gamma$ in the co-moving frame [Fig. 1]. Similarly, the lattice spacing along the $z'$-direction is also increased. For simplicity, in this study, we fix the crystal geometry in the co-moving frame. Specifically, we assume that in the co-moving coordinates, the geometry of the inclusions is spherical with a radius *R* and that the inclusions are arranged in a cubic lattice with a period *a*. Thus, in the lab frame coordinates, the inclusions appear slightly ellipsoidal and are arranged in an orthorhombic lattice. Figure 1b illustrates the details of the unit cell geometry as observed in the lab frame coordinates. The path traversed by the inclusion over time is represented with a horizontal line shaded in orange. Unlike other configurations studied in the literature [21-26], the region of the unit cell subjected to spacetime modulation is localized near this line segment. Consequently, this geometry may offer practical advantages for implementation compared to geometries where the material parameters need to be modulated throughout the entire space.

### B. *Generalized Clausius-Mossotti homogenization*

Next, we generalize the standard Clausius-Mossotti approach [28, 29, 30] to spacetime crystals. The key idea is to model a generic spherical particle in the co-moving frame as a superposition of an electric dipole ($\mathbf{p}'$) and a magnetic dipole ($\mathbf{m}'$). The electric dipole moment and the magnetic dipole moment are related to the local electric and magnetic fields through a generalized polarizability tensor $\boldsymbol{\alpha}$ as follows:

$$\begin{pmatrix} \mathbf{p}'/\varepsilon_0 \\ \mathbf{m}' \end{pmatrix} = \boldsymbol{\alpha} \cdot \begin{pmatrix} \mathbf{E}'_{\text{loc}} \\ \mathbf{H}'_{\text{loc}} \end{pmatrix}. \tag{6}$$

The local fields ($\mathbf{E}'_{\text{loc}}, \mathbf{H}'_{\text{loc}}$) are evaluated at the particle position with the self-field excluded. The generalized polarizability tensor has units of volume and accounts for the bianisotropic response of the inclusions in the co-moving frame coordinates. Using a quasi-static approximation, we show in Appendix A that the polarizability can be written in terms of the spherical inclusions parameters as [Eq. (A5)]:

$$\boldsymbol{\alpha} = \alpha_0 \left( \mathbf{M}' + 2\mathbf{M}_0 \right)^{-1} \cdot \left( \mathbf{M}' - \mathbf{M}_0 \right), \tag{7}$$

with $\alpha_0 = 4\pi R^3$. Here, $\mathbf{M}_0 = \begin{pmatrix} \varepsilon_0 \mathbf{1}_{3\times 3} & 0 \\ 0 & \mu_0 \mathbf{1}_{3\times 3} \end{pmatrix}$ is the material matrix of the vacuum with $\varepsilon_0, \mu_0$ the permittivity and permeability of vacuum.

The effective response in the co-moving frame can now be determined by relating the local fields $\mathbf{E}'_{\text{loc}}, \mathbf{H}'_{\text{loc}}$ with the averaged (macroscopic) fields $\langle \mathbf{E}' \rangle, \langle \mathbf{H}' \rangle$. From the Lorentz-Lorenz formalism, it is known that in the quasi-static limit [28, 32]:

$$\mathbf{E}'_{\text{loc}} = \langle \mathbf{E}' \rangle + \mathbf{C}_{\text{int}} \cdot \frac{\mathbf{p}'}{\varepsilon_0}, \qquad \mathbf{H}'_{\text{loc}} = \langle \mathbf{H}' \rangle + \mathbf{C}_{\text{int}} \cdot \mathbf{m}'. \tag{8}$$

Here, $\mathbf{C}_{\text{int}}$ is an interaction tensor that accounts for the field created by all the particles in the crystal on the position of a generic particle. For a cubic lattice, the interaction tensor reduces to a scalar $\mathbf{C}_{\text{int}} = \mathbf{1}_{3\times 3}/(3a^3)$ [28, 32]. Hence, it follows that,

$$\begin{pmatrix} \mathbf{E}'_{\text{loc}} \\ \mathbf{H}'_{\text{loc}} \end{pmatrix} = \begin{pmatrix} \langle \mathbf{E}' \rangle \\ \langle \mathbf{H}' \rangle \end{pmatrix} + \frac{1}{3a^3} \begin{pmatrix} \mathbf{p}'/\varepsilon_0 \\ \mathbf{m}' \end{pmatrix}. \tag{9}$$

By combining the above formula with Eq. (6) it becomes possible to express the dipole moments in terms of the macroscopic fields:

$$\begin{pmatrix} \mathbf{p}'/\varepsilon_0 \\ \mathbf{m}' \end{pmatrix} = \left( \mathbf{1}_{6\times 6} - \frac{\boldsymbol{\alpha}}{3a^3} \right)^{-1} \cdot \boldsymbol{\alpha} \cdot \begin{pmatrix} \langle \mathbf{E}' \rangle \\ \langle \mathbf{H}' \rangle \end{pmatrix}. \tag{10}$$

The macroscopic fields $\langle \mathbf{D}' \rangle, \langle \mathbf{B}' \rangle$ are determined by $\langle \mathbf{D}' \rangle = \varepsilon_0 \langle \mathbf{E}' \rangle + \frac{1}{a^3}\mathbf{p}'$ and $\langle \mathbf{B}' \rangle = \mu_0 \langle \mathbf{H}' \rangle + \frac{\mu_0}{a^3}\mathbf{m}'$. Thus, we conclude that the effective material matrix in the co-moving frame is:

$$\begin{aligned} \mathbf{M}'_{\text{ef}} &= \mathbf{M}_0 \cdot \left[ \mathbf{1}_{6\times 6} + \left( \mathbf{1}_{6\times 6} - \frac{\boldsymbol{\alpha}}{3a^3} \right)^{-1} \cdot \frac{\boldsymbol{\alpha}}{a^3} \right] \\ &= \mathbf{M}_0 \cdot \left[ \mathbf{1}_{6\times 6} + \left( a^3 \boldsymbol{\alpha}^{-1} - \frac{1}{3a^3} \mathbf{1}_{6\times 6} \right)^{-1} \right]. \end{aligned} \tag{11}$$

The second identity assumes that $\boldsymbol{\alpha}$ has an inverse. The previous analysis mirrors the standard Clausius-Mossotti homogenization, here extended to bianisotropic particles. Finally, the effective response in the lab frame can be found with an inverse Lorentz transformation (compare with Eq. (4)):

$$\mathbf{M}_{\text{ef}} \equiv \begin{pmatrix} \varepsilon_0 \overline{\varepsilon_{\text{ef}}} & \frac{1}{c}\overline{\xi_{\text{ef}}} \\ \frac{1}{c}\overline{\zeta_{\text{ef}}} & \mu_0 \overline{\mu_{\text{ef}}} \end{pmatrix} = \left[ -\frac{1}{c^2}\mathbf{V} + \mathbf{A} \cdot \mathbf{M}'_{\text{ef}} \right] \cdot \left[ \mathbf{A} - \mathbf{V} \cdot \mathbf{M}'_{\text{ef}} \right]^{-1}. \tag{12}$$

In summary, the quasi-static response in the lab frame is given by Eq. (12) with $\mathbf{M}'_{\text{ef}}$ determined by the co-moving frame polarizability $\boldsymbol{\alpha} = \alpha_0 (\mathbf{M}' + 2\mathbf{M}_0)^{-1} \cdot (\mathbf{M}' - \mathbf{M}_0)$ through Eq. (11), and $\mathbf{M}'$ is provided by Eq. (4). Note that $\mathbf{M}_{\text{ef}}$ is a 6×6 real-valued matrix in the quasi-static limit. In general, the effective response in the lab frame is bianisotropic so that the magneto-electric tensors $\overline{\xi_{\text{ef}}}, \overline{\zeta_{\text{ef}}}$ are nontrivial.

*C. Taylor series approximation of the effective material parameters*

To gain some intuition on the effective response of the spacetime crystal, next we perform a Taylor series expansion of $\mathbf{M}_{\text{ef}}$. To achieve this, we express the relative permittivity and relative permeability of the spheroid inclusions as a small perturbation relative to the background material: $\overline{\varepsilon} = \mathbf{1}_{3\times 3} + \delta\boldsymbol{\varepsilon}$ and $\overline{\mu} = \mathbf{1}_{3\times 3} + \delta\boldsymbol{\mu}$. A first-order Taylor series of $\mathbf{M}_{ef}$ in the small parameter $v/c$ (terms on the order of $(v/c)^2$ are discarded), followed by a second-order Taylor series with respect to the small parameters $\delta\boldsymbol{\varepsilon}$, $\delta\boldsymbol{\mu}$, yields the approximate formulas:

$$\overline{\varepsilon}_{ef} \approx \mathbf{1}_{3\times 3} + f_V \delta\boldsymbol{\varepsilon} - \frac{f_V(1-f_V)}{3}\delta\boldsymbol{\varepsilon}\cdot\delta\boldsymbol{\varepsilon}, \tag{13a}$$

$$\overline{\mu}_{ef} \approx \mathbf{1}_{3\times 3} + f_V \delta\boldsymbol{\mu} - \frac{f_V(1-f_V)}{3}\delta\boldsymbol{\mu}\cdot\delta\boldsymbol{\mu}, \tag{13b}$$

$$\overline{\xi}_{ef} = \overline{\zeta}_{ef}^T \approx -\frac{f_V(1-f_V)}{3}\delta\boldsymbol{\varepsilon}\cdot\left(\frac{\mathbf{v}}{c}\times\mathbf{1}_{3\times 3}\right)\cdot\delta\boldsymbol{\mu}, \tag{13c}$$

where $f_V = \frac{4\pi R^3}{3a^3}$ is the volume fraction of the inclusions in the co-moving frame. The derivation is omitted for conciseness. As seen, the effective permittivity and the effective permeability are insensitive to the time modulation when the terms on the order $(v/c)^2$ are discarded. In contrast, the magneto-electric tensors scale proportionally to the modulation speed. Furthermore, analogous to Ref. [21, 22], a nontrivial bianisotropic response can occur only when both the permittivity and permeability tensors are simultaneously modulated in space and in time. This property remains valid even for strong modulations of the material parameters and large $v/c$.

For an isotropic perturbation of the permittivity and permeability ($\delta\boldsymbol{\varepsilon} = \delta\varepsilon\mathbf{1}_{3\times 3}$ and $\delta\boldsymbol{\mu} = \delta\mu\mathbf{1}_{3\times 3}$) it is evident that the nonreciprocal coupling is consistent with a moving medium coupling [20, 21, 25]. Specifically, the effective material parameters take the form $\overline{\varepsilon}_{ef} \approx \varepsilon_{ef}\mathbf{1}_{3\times 3}$, $\overline{\mu}_{ef} \approx \mu_{ef}\mathbf{1}_{3\times 3}$ and $\overline{\xi}_{ef} = \overline{\zeta}_{ef}^T = -\xi_{ef}\left(\frac{\mathbf{v}}{c}\times\mathbf{1}_{3\times 3}\right)$ with

$$\varepsilon_{ef} \approx 1 + f_V \delta\varepsilon - \frac{f_V(1-f_V)}{3}(\delta\varepsilon)^2, \qquad \mu_{ef} \approx 1 + f_V \delta\mu - \frac{f_V(1-f_V)}{3}(\delta\mu)^2 \qquad \text{and}$$

$\xi_{ef} = \frac{f_V(1-f_V)}{3}\delta\varepsilon\delta\mu$. To illustrate the range of validity of the Taylor expansion, we plot in Fig. 2 the effective parameters as a function of modulation speed $v$ for $\delta_\varepsilon = \delta_\mu = 0.1$. The solid lines represent the exact homogenization result [Eq. (12)] and the dashed lines represent the Taylor approximation for a weak modulation [Eq. (13)]. As seen, the Taylor approximation captures very well the effective response even for fairly large modulation velocities.

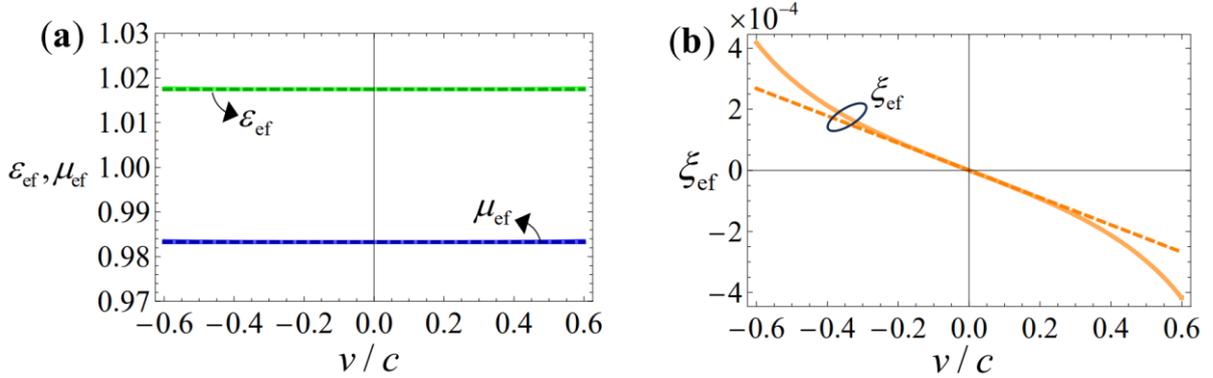

**Fig. 2** Components $\varepsilon_{ef}$, $\mu_{ef}$ and $\xi_{ef}$ of the effective material response as a function of the modulation speed for $\delta_\varepsilon = \delta_\mu = 0.1$. The solid lines represent the full homogenization result [Eq. (12)], whereas the dashed lines represent the effective parameters obtained with the approximate analytical formulas for a weak modulation [Eq. (13)]. The radius of the spheroid inclusions is $R = 0.35a$ and the modulation speed is $\mathbf{v} = v\hat{\mathbf{z}}$.

It is possible to engineer a wide range of nonreciprocal responses by controlling $\delta\boldsymbol{\varepsilon}$ and $\delta\boldsymbol{\mu}$, and most notably the anisotropic response of material tensors. This will be discussed in detail in the following subsection. However, it should be noted $\overline{\overline{\xi}}_{ef}$ has a vanishing determinant because $\det\left(\frac{\mathbf{v}}{c} \times \mathbf{1}_{3\times3}\right) = 0$. Thus, the most general class of nonreciprocal responses that can be engineered using spacetime modulated crystals are limited by the constraint $\det\left(\overline{\overline{\xi}}_{\text{ef}}\right) = 0$. Numerical simulations (not shown here) indicate that the constraint holds true even for strong modulations of the material response and large $v/c$.

*D. Moving-Tellegen response*

In order to illustrate how the anisotropy of the spheroid inclusions tailors the effective response of the spacetime crystal, next we consider permittivity and permeability tensors with the following structure:

$$\begin{aligned}
\overline{\overline{\varepsilon}} &= \varepsilon_1 \hat{\mathbf{e}}_1 \otimes \hat{\mathbf{e}}_1 + \varepsilon_2 \hat{\mathbf{e}}_2 \otimes \hat{\mathbf{e}}_2 + \varepsilon_v \hat{\mathbf{v}} \otimes \hat{\mathbf{v}}, \\
\overline{\overline{\mu}} &= \mu_1 \hat{\mathbf{u}}_1 \otimes \hat{\mathbf{u}}_1 + \mu_2 \hat{\mathbf{u}}_2 \otimes \hat{\mathbf{u}}_2 + \mu_v \hat{\mathbf{v}} \otimes \hat{\mathbf{v}}.
\end{aligned} \quad (14)$$

Here, $\hat{\mathbf{e}}_1, \hat{\mathbf{e}}_2$ and $\hat{\mathbf{u}}_1, \hat{\mathbf{u}}_2$ are the principal axes of the permittivity and permeability in the *xoy* plane, respectively, and $\hat{\mathbf{v}} = \hat{\mathbf{z}}$ is the unity vector that gives the direction of the modulation speed $\mathbf{v} = v\hat{\mathbf{v}}$. It is assumed that $\hat{\mathbf{e}}_1 \times \hat{\mathbf{e}}_2 = \hat{\mathbf{u}}_1 \times \hat{\mathbf{u}}_2 = \hat{\mathbf{v}}$. In general, the permittivity and permeability tensors do not share the same principal axes in the *xoy* plane (see Fig. 3a). The angular offset between the permeability and permittivity principal axes is denoted by $\theta$. The parameters $\varepsilon_1, \varepsilon_2, \varepsilon_v$ and $\mu_1, \mu_2, \mu_v$ determine the eigenvalues of the permittivity and permeability tensors. The considered structure of the material response is inspired by our previous work [22].

Next, we present a Taylor series approximation for the analytical expressions of the constitutive parameters as a function of the modulation speed. Different from Eq. (13), here we do not make any assumptions about the strength of $v/c$, and only suppose that the perturbations $\delta \boldsymbol{\varepsilon}$ and $\delta \boldsymbol{\mu}$ with respect to the background material are weak. Furthermore, for simplicity we assume that:

$$\varepsilon_1 = 1 + \delta_\varepsilon, \qquad \varepsilon_2 = 1 - \delta_\varepsilon, \qquad \mu_1 = 1 + \delta_\mu, \qquad \mu_2 = 1 - \delta_\mu, \qquad \varepsilon_v = \mu_v = 1. \quad (15)$$

The second order Taylor series approximation of Eq. (12) with respect to the small parameters $\delta_\varepsilon, \delta_\mu$ yields:

$$\overline{\overline{\varepsilon}}_{\text{ef}} = \mathbf{1} + f_V \delta_\varepsilon \left( \hat{\mathbf{e}}_1 \otimes \hat{\mathbf{e}}_1 - \hat{\mathbf{e}}_2 \otimes \hat{\mathbf{e}}_2 \right) + \Delta_\varepsilon \left( \mathbf{1} - \hat{\mathbf{v}} \otimes \hat{\mathbf{v}} \right),$$
$$\overline{\overline{\mu}}_{\text{ef}} = \mathbf{1} + f_V \delta_\mu \left( \hat{\mathbf{u}}_1 \otimes \hat{\mathbf{u}}_1 - \hat{\mathbf{u}}_2 \otimes \hat{\mathbf{u}}_2 \right) + \Delta_\mu \left( \mathbf{1} - \hat{\mathbf{v}} \otimes \hat{\mathbf{v}} \right), \quad (16\text{a})$$
$$\overline{\overline{\xi}}_{\text{ef}} = \overline{\overline{\zeta}}_{\text{ef}}^{\text{T}} = \xi \left[ \left( \mathbf{1} - \hat{\mathbf{v}} \otimes \hat{\mathbf{v}} \right) \sin(2\theta) - \hat{\mathbf{v}} \times \mathbf{1} \cos(2\theta) \right],$$

with

$$\Delta_\varepsilon = -\frac{\delta_\varepsilon^2}{3} f_V (1 - f_V) \frac{1 - 2\beta^2}{1 - \beta^2}$$
$$\Delta_\mu = -\frac{\delta_\mu^2}{3} f_V (1 - f_V) \frac{1 - 2\beta^2}{1 - \beta^2}, \quad (16\text{b})$$
$$\xi = -\frac{\delta_\mu \delta_\varepsilon}{3} f_V (1 - f_V) \frac{\beta}{1 - \beta^2}$$

and $\beta = v/c$. When the terms involving $\beta^2$ are dropped, the above formulas reduce to Eq. (13).

The effective magneto-electric tensor is proportional to the product of the modulation strengths $\delta_\mu, \delta_\varepsilon$, and depends on the offset $\theta$. The nonreciprocal coupling in Eq. (16) is known as a "moving-Tellegen" response [22], and is qualitatively similar to the one reported for layered crystals in Ref. [56]. In particular, for aligned principal axes ($\theta = n\frac{\pi}{2}, n = 0,1,..$) the spacetime crystal exhibits a moving medium coupling response described by an anti-symmetric tensor of the type $\overline{\overline{\xi}}_{\text{ef}} \sim \hat{\mathbf{v}} \times \mathbf{1}$, whereas for an axis offset such that $\theta = \frac{\pi}{4} + n\frac{\pi}{2}$ ($n = 0,1,..$), the crystal exhibits a Tellegen response described by a symmetric tensor of the type: $\overline{\overline{\xi}}_{\text{ef}} \sim \mathbf{1} - \hat{\mathbf{v}} \otimes \hat{\mathbf{v}}$. The effective permittivity and permeability tensors do not depend explicitly on the relative angle $\theta$ as they do not include a contribution that simultaneously involves the electric and magnetic responses.

It is useful to write the tensors in Eqs. (16) with respect to a fixed Cartesian coordinate basis. For simplicity, we suppose that the principal axes of the permeability are aligned with the $x$ and $y$ directions, $\hat{\mathbf{u}}_1 \equiv \hat{\mathbf{x}}$, $\hat{\mathbf{u}}_2 \equiv \hat{\mathbf{y}}$, and we take $\hat{\mathbf{u}}_3 \equiv \hat{\mathbf{z}}$. Then, we may write the relevant tensors as $\bar{\bar{\varepsilon}}_{ef} = [\varepsilon_{ef,ij}]_{i,j=1,...3}$, $\bar{\bar{\mu}}_{ef} = [\mu_{ef,ij}]_{i,j=1,...3}$, and $\bar{\bar{\xi}}_{ef} = [\xi_{ef,ij}]_{i,j=1,...3}$. Explicit formulas for the matrix elements are given in Appendix B.

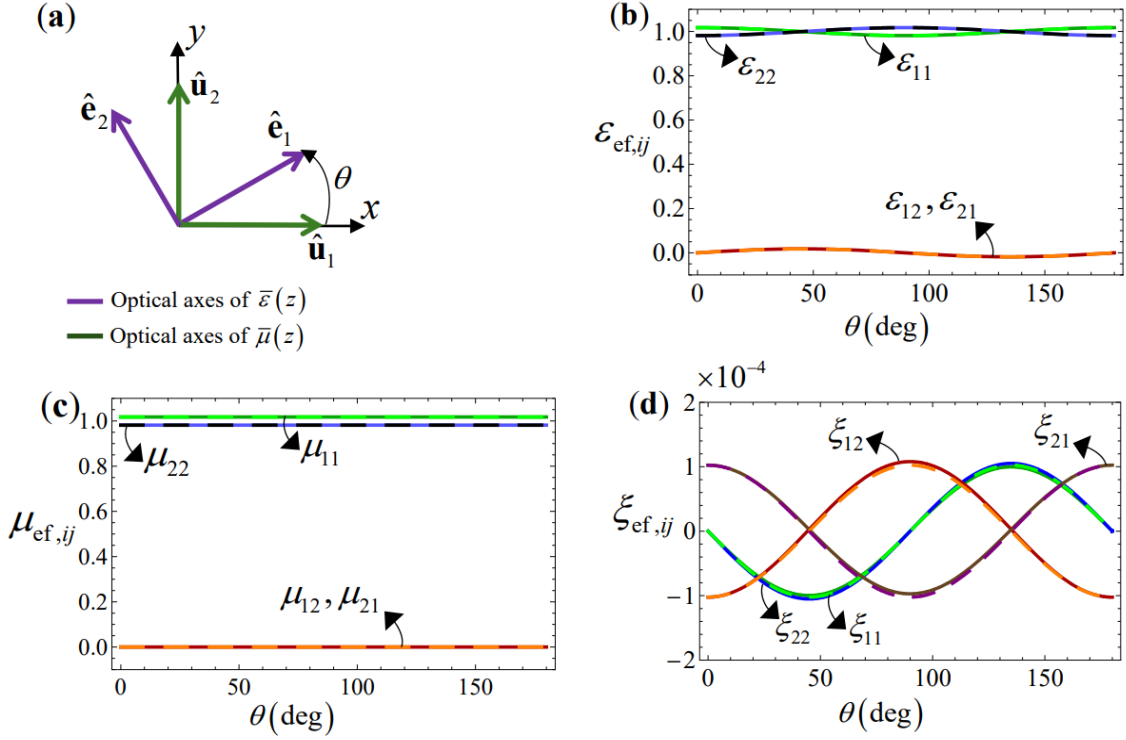

**Fig. 3 (a)** Optical axes of the permittivity and permeability tensors, $\bar{\bar{\varepsilon}}(z)$ and $\bar{\bar{\mu}}(z)$, respectively, in the lab frame coordinates. The offset angle is denoted by $\theta$. **(b)-(d)** Components of the tensors $\bar{\bar{\varepsilon}}_{ef}, \bar{\bar{\mu}}_{ef}, \bar{\bar{\xi}}_{ef}$, as a function of the angle $\theta$, for a modulation with $\delta_\varepsilon = \delta_\mu = 0.1$. The solid lines represent the full homogenization result [Eq. (12)], whereas the dashed lines rely on the weak modulation approximation [Eqs. (B1)-(B3)]. The modulation speed is $v = 0.2c$, and the radius of the spheroid inclusions is $R = 0.35a$. The indices *i,j* label the elements of each tensor. **(b)** Components of the effective permittivity tensor. **(c)** Components of the effective permeability tensor. **(d)** Components of the magneto-electric tensor.

The components of the effective tensors are depicted in Fig. 3b-d as a function of $\theta$, for $\delta_\varepsilon = \delta_\mu = 0.1$ and modulation speed $v = 0.2c$. The solid lines represent the exact homogenization result [Eq. (12)], whereas the dashed lines are obtained using the approximate analytical formulas for a weak modulation [Eq. (16) or equivalently Eqs. (B1)- (B3)]. As seen, for $\delta_\varepsilon = \delta_\mu = 0.1$, the approximate analytical formulas match very well the full homogenization formulas. Note that in this example $\delta\boldsymbol{\mu}$ is kept fixed, whereas the principal axes of $\delta\boldsymbol{\varepsilon}$ depend on the angle $\theta$. This justifies why the elements of the effective permeability tensor are practically independent of $\theta$. The tensor $\overline{\overline{\mu}}_{\text{ef}}$ is diagonal because the principal axes of $\delta\boldsymbol{\mu}$ are aligned with the coordinate axis. Interestingly, Fig. 3d reveals that by controlling the alignment of the principal axes of $\delta\boldsymbol{\mu}$ and $\delta\boldsymbol{\varepsilon}$ it is possible to continuously transform a moving medium type response ($\theta = 0º, 90º, 180º$, where $\xi_{\text{ef},11} = \xi_{\text{ef},22} = 0$ and $\xi_{\text{ef},12} = -\xi_{\text{ef},21}$) into a Tellegen-type response ($\theta = 45º, 135º$, where $\xi_{\text{ef},11} = \xi_{\text{ef},22}$ and $\xi_{\text{ef},12} = \xi_{\text{ef},21} = 0$).

Next, we focus on the case $\theta = 45º$, corresponding to a transverse Tellegen response described the symmetric real-valued tensor $\overline{\overline{\xi}}_{\text{ef}}\big|_{\theta=45º} = \kappa(\mathbf{1} - \hat{\mathbf{v}} \otimes \hat{\mathbf{v}})$, with $\kappa = -\dfrac{\delta_\mu \delta_\varepsilon}{3} f_V (1-f_V) \dfrac{\beta}{1-\beta^2}$ the Tellegen parameter. Figures 4a-c show the relevant components of the effective permittivity, permeability and magneto-electric coupling as a function of the modulation speed, for the same structural parameters as in Fig. 3 (weak modulation, $\delta_\varepsilon = \delta_\mu = 0.1$). As seen, the effective permittivity and permeability are weakly sensitive to variations in $v$. In contrast, the Tellegen parameter varies linearly with $v$ for small velocities and even more strongly for large velocities. It is relevant to note that the Taylor approximation works very well even for relativistic modulation speeds. For $v = 0.8c$, the Tellegen parameter can be as large as $\kappa = -0.0011$.

It is possible to obtain larger values of the magnitude of the Tellegen parameter by increasing the modulation strength or the volume fraction of the inclusions. For instance, with a strong modulation ($\delta_\varepsilon = \delta_\mu = 0.5$), the Tellegen parameter for $v = 0.8c$ becomes roughly 25 times larger ($\kappa = -0.027$), as seen in Fig. 4d. This plot also shows for a strong modulation the Taylor approximation becomes less accurate, necessitating the use of exact formulas [Eq. (16)]. Furthermore, under strong modulation, some structural anisotropy emerges in the magneto-electric tensor ($\xi_{ef,11} \neq \xi_{ef,22}$).

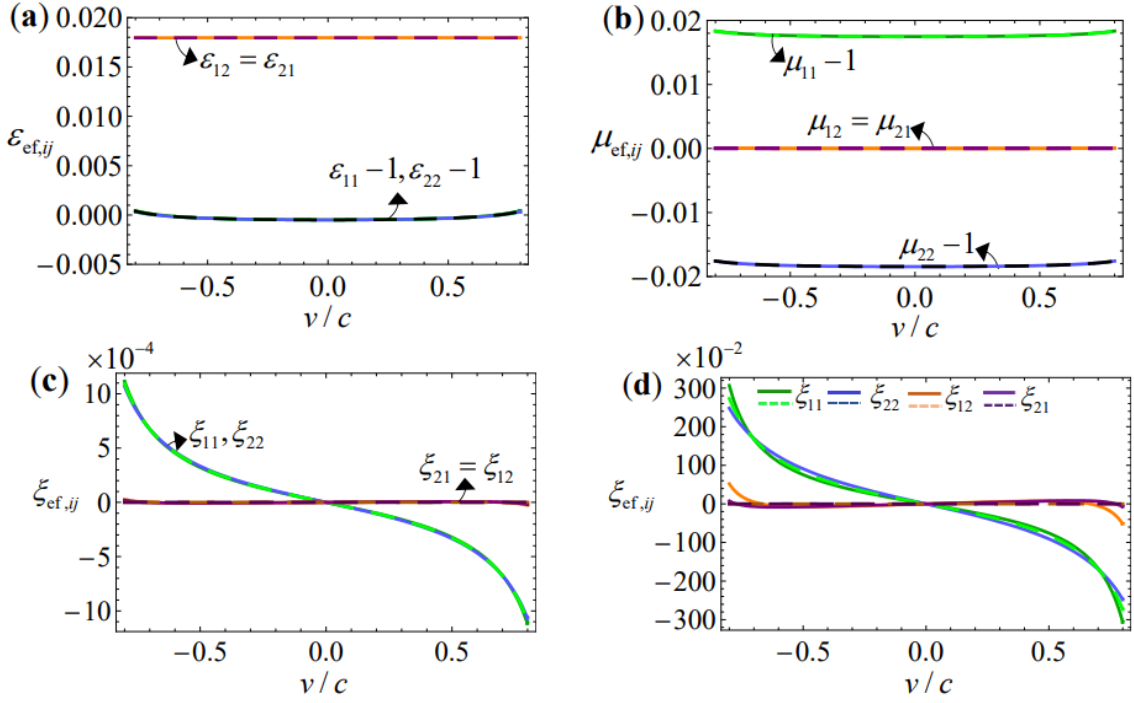

**Fig. 4 (a)-(d)** Components of the tensors $\overline{\overline{\varepsilon}}_{ef}, \overline{\overline{\mu}}_{ef}, \overline{\overline{\xi}}_{ef}$ for a spacetime crystal with an effective Tellegen coupling [$\theta = 45°$], as a function of the modulation speed $v$. The structural parameters are $R = 0.35a$ and $\delta_\varepsilon = \delta_\mu = 0.1$ (weak modulation). **(d)** Similar to (c) but for a strong modulation $\delta_\varepsilon = \delta_\mu = 0.5$. In all plots, the solid lines are calculated with the full homogenization result [Eq. (12)], and the dashed lines with the Taylor approximation [Eqs. (B1)-(B3) with $\theta = 45°$].

### E. Isotropic Magneto-electric Coupling

An ideal Tellegen material is characterized by an isotropic nonreciprocal response, where both $\overline{\overline{\xi}}_{ef}, \overline{\overline{\zeta}}_{ef}$ are scalars and identical [36]. This means that an electric excitation of the material induces both electric and magnetic dipoles, each aligned with the applied electric field. Similarly, a magnetic excitation induces both electric and magnetic dipoles aligned with the applied magnetic field. Tellegen envisioned that such a material response could be realized by randomly mixing nanoparticles with permanent and parallel electric and magnetic dipoles, with the orientation of the nanoparticles being sensitive to the applied external fields.

As $\overline{\xi}_{ef}, \overline{\zeta}_{ef}$ are identical, the Tellegen medium has a broken time-reversal symmetry. Unlike other established solutions to break reciprocity, which inherently provide anisotropic responses due to the presence of some bias, the Tellegen proposal offers an isotropic magneto-electric coupling. This counter-intuitive property has been the subject of criticism and dispute [57-59].

Evidently, due to the constraint $\det(\overline{\xi}_{ef}) = 0$, our proposal based on spacetime modulations cannot provide an isotropic Tellegen coupling. In fact, the modulation velocity **v** determines a special direction in space, resulting in a structure for $\overline{\xi}_{ef}$ that is incompatible with an isotropic response. Therefore, to achieve an isotropic response, we need a form of spacetime modulation that does not single out any particular direction in space.

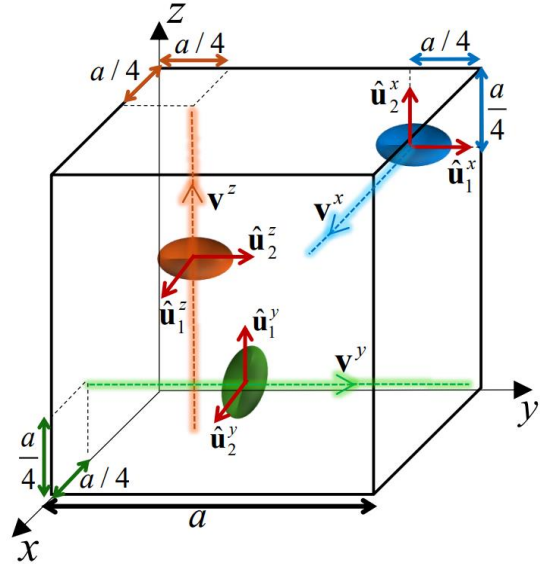

**Fig. 5** Geometry of the unit cell of a spacetime crystal that provides an isotropic magneto-electric coupling. The unit cell is formed by three spheroid inclusions, each subject to a modulation velocity $\mathbf{v}^i$, (i=x,y,z), with $|\mathbf{v}^i| = v = const$. The figure also shows the principal axes of the permeability tensor of each inclusion $\hat{\mathbf{u}}_1^i$ and $\hat{\mathbf{u}}_2^i$ with $\hat{\mathbf{u}}_1^i \times \hat{\mathbf{u}}_2^i = \hat{\mathbf{v}}^i$. The principal axes of the permittivity tensor $\hat{\mathbf{e}}_1^i$ and $\hat{\mathbf{e}}_2^i$ are rotated by 45° with respect to the axes of the permeability tensor (not shown).

We propose to implement an isotropic Tellegen coupling based on the modulation sketched in Fig. 5. The unit cell contains three spheroid inclusions, each subjected to a modulation velocity along one of the coordinate axis, $\mathbf{v}^x = v\hat{\mathbf{x}}$, $\mathbf{v}^y = v\hat{\mathbf{y}}$ or $\mathbf{v}^z = v\hat{\mathbf{z}}$, respectively. The trajectories of the inclusions are designed to maintain maximal distance from each other to minimize electromagnetic coupling: $p_X = \left(vt, \frac{3a}{4}, \frac{3a}{4}\right)$, $p_Y = \left(\frac{3a}{4}, vt, \frac{a}{4}\right)$ and $p_Z = \left(\frac{a}{4}, \frac{a}{4}, vt\right)$.

Assuming weak coupling between different inclusions (specifically, each set of inclusions should perceive the remaining inclusions as an effective medium [60]), the effective response of the crystal with three particles in the unit cell ($\mathbf{M}_{\text{ef}}$) can be found from the effective responses of crystals with a single particle per cell ($\mathbf{M}_{\text{ef}}^{(i)}$, $i=x,y,z$), as follows:

$$\mathbf{M}_{\text{ef}} = \mathbf{M}_0 + \sum_{i=x,y,z} \left(\mathbf{M}_{\text{ef}}^{(i)} - \mathbf{M}_0\right). \tag{17}$$

The tensors $\mathbf{M}_{\text{ef}}^{(i)}$ can be calculated using Eq. (16). We assume that the $i$-th subset of inclusions is described by:

$$\begin{aligned} \bar{\varepsilon}^{(i)} &= \varepsilon_1 \hat{\mathbf{e}}_1^i \otimes \hat{\mathbf{e}}_1^i + \varepsilon_2 \hat{\mathbf{e}}_2^i \otimes \hat{\mathbf{e}}_2^i + \hat{\mathbf{v}}^i \otimes \hat{\mathbf{v}}^i, \\ \bar{\mu}^{(i)} &= \mu_1 \hat{\mathbf{u}}_1^i \otimes \hat{\mathbf{u}}_1^i + \mu_2 \hat{\mathbf{u}}_2^i \otimes \hat{\mathbf{u}}_2^i + \hat{\mathbf{v}}^i \otimes \hat{\mathbf{v}}^i. \end{aligned} \tag{18}$$

Here, $\varepsilon_1, \varepsilon_2, \mu_1, \mu_2$ are defined as in Eq. (15) and are independent of the inclusion subset ($i=x,y,z$). The radius $R$ is also independent of the inclusion subset. The orientation of the principal axes ($\hat{\mathbf{u}}_1^i, \hat{\mathbf{u}}_2^i$) of the permeability tensor $\bar{\mu}^{(i)}$ is shown in Fig. 5, ensuring that $\hat{\mathbf{u}}_1^i \times \hat{\mathbf{u}}_2^i = \hat{\mathbf{v}}^i$. Furthermore, the orientation of the principal axes of the permittivity is governed by $\hat{\mathbf{e}}_1^i = \cos\theta \hat{\mathbf{u}}_1^i + \sin\theta \hat{\mathbf{u}}_2^i$ and $\hat{\mathbf{e}}_2^i = -\sin\theta \hat{\mathbf{u}}_1^i + \cos\theta \hat{\mathbf{u}}_2^i$ with $\theta = 45°$, to guarantee that the contribution of each sub-lattice to the effective magneto-electric coupling tensor is a diagonal matrix.

Formula (17) can be expanded in terms of the permittivity, permeability and magneto-electric responses as:

$$\begin{aligned}\overline{\overline{\varepsilon}}_{\text{ef}} &= \overline{\overline{\varepsilon}}_{\text{ef}}^{x} + \overline{\overline{\varepsilon}}_{\text{ef}}^{y} + \overline{\overline{\varepsilon}}_{\text{ef}}^{z} - 2\mathbf{1}_{3\times3} \\ \overline{\overline{\mu}}_{\text{ef}} &= \overline{\overline{\mu}}_{\text{ef}}^{x} + \overline{\overline{\mu}}_{\text{ef}}^{y} + \overline{\overline{\mu}}_{\text{ef}}^{z} - 2\mathbf{1}_{3\times3} \\ \overline{\overline{\xi}}_{\text{ef}} &= \overline{\overline{\xi}}_{\text{ef}}^{x} + \overline{\overline{\xi}}_{\text{ef}}^{y} + \overline{\overline{\xi}}_{\text{ef}}^{z}\end{aligned} \quad (19)$$

It is straightforward to check that for the described geometry and for a weak modulation, one has $\overline{\overline{\xi}}_{\text{ef}}^{(i)} = \kappa\left(\mathbf{1}_{3\times3} - \hat{\mathbf{v}}^{i} \otimes \hat{\mathbf{v}}^{i}\right)$, similar to the result described in the subsection II.D. Thereby, it follows that under the described assumptions the magneto-electric tensor reduces to a scalar:

$$\overline{\overline{\xi}}_{\text{ef}} = \kappa_{\text{ef}}\mathbf{1}_{3\times3}, \quad \text{with} \quad \kappa_{\text{ef}} = 2\kappa = -\frac{2}{3}\delta_{\mu}\delta_{\varepsilon}f_{V}\left(1-f_{V}\right)\frac{\beta}{1-\beta^{2}}. \quad (20)$$

Hence, the proposed modulation can indeed provide an isotropic magneto-electric nonreciprocal response, analogous to the solution envisioned by Bernard Tellegen. It is straightforward to check that the effective medium is also characterized by an isotropic magnetic response $\overline{\overline{\mu}}_{\text{ef}} = \left(1 + 2\Delta_{\mu}\right)\mathbf{1}_{3\times3}$. On the other hand, $\overline{\overline{\varepsilon}}_{\text{ef}}$ does not reduce to a scalar, and thereby the electric response exhibits some anisotropy (not shown). This residual anisotropy in the electric response could be compensated by adding an anisotropic static dielectric particle to the unit cell. The amplitude of the Tellegen parameter $\kappa_{\text{ef}}$ is two times larger than the amplitude of the Tellegen parameter for a crystal with a single particle per cell.

### III. Generalized Minkowskian spacetime crystals

Minkowskian spacetime crystals were recently introduced as time-variant systems described by constitutive relations that are observer independent [55]. Next, we extend this concept and prove that the most general class of materials invariant under a Lorentz boost along a fixed spatial direction is formed by pseudo-uniaxial crystals. We elucidate how to take advantage of the unique properties of such time-variant systems to characterize in an expedite manner their band structure. Furthermore, we use the developed formalism to validate our Clausius-Mossotti homogenization theory.

### A. Media invariant under Lorentz boosts with a fixed direction

Our previous works [55, 61] exploit the fact that media with $\varepsilon\mu = n^2 = 1$ stay invariant under an arbitrary Lorentz transformation [62]. This invariance allows one to connect the response of an isorefractive spacetime crystal with that of the corresponding static crystal (deformed according to the Lorentz-Fitzgerald contraction) without altering the constitutive relations. However, requiring the materials to be invariant under arbitrary transformations is overly restrictive. Practically, the same connection can be established as long as the constitutive relations of the materials remain unchanged under the specific Lorentz boost that renders the system time-invariant. In other words, it suffices to ensure that the co-moving frame material matrix $\mathbf{M}'$, given by Eq. (4), coincides with the lab material matrix $\mathbf{M}$ for a Lorentz boost directed along the modulation velocity $\mathbf{v}$.

In Appendix C, we demonstrate that a general class of dielectric media invariant under Lorentz boosts aligned with the z-direction ($\mathbf{v} = v\hat{\mathbf{z}}$) have electric and magnetic responses of the type:

$$\overline{\varepsilon} = \begin{pmatrix} \boldsymbol{\varepsilon}_\perp & \mathbf{0}_{2\times 1} \\ \mathbf{0}_{1\times 2} & \varepsilon_{zz} \end{pmatrix}, \qquad \overline{\mu} = \delta\mu_{zz}\hat{\mathbf{z}}\otimes\hat{\mathbf{z}} + \left[\mathbf{1}_{3\times 3} - \hat{\mathbf{z}}\times(\overline{\varepsilon}-\mathbf{1}_{3\times 3})\times\hat{\mathbf{z}}\right]^{-1}. \tag{21}$$

In the above, $\varepsilon_{zz}, \delta\mu_{zz}$ are arbitrary scalars and $\boldsymbol{\varepsilon}_\perp$ is an arbitrary 2×2 matrix. For reciprocal dielectrics ($\boldsymbol{\varepsilon}_\perp$ is a real symmetric matrix), one can always pick a coordinate system such that the permittivity tensor is diagonal. In such a case, Eq. (21) reduces to (with $\mu_{zz} = 1+\delta\mu_{zz}$):

$$\bar{\varepsilon} = \begin{pmatrix} \varepsilon_{xx} & 0 & 0 \\ 0 & \varepsilon_{yy} & 0 \\ 0 & 0 & \varepsilon_{zz} \end{pmatrix}, \quad \bar{\mu} = \begin{pmatrix} 1/\varepsilon_{yy} & 0 & 0 \\ 0 & 1/\varepsilon_{xx} & 0 \\ 0 & 0 & \mu_{zz} \end{pmatrix}. \tag{22}$$

The electromagnetic response of such media is invariant under any Lorentz boost directed along the $z$-direction ($\mathbf{M}' = \mathbf{M}$). This analysis extends Minkowskian isotropic materials to systems with lower symmetry, potentially leading to more practical designs.

It is interesting to characterize the plane wave dispersion in media with the constitutive relations (22). There are two independent polarizations with dispersions,

$$\frac{\varepsilon_{xx}}{\varepsilon_{zz}} k_x^2 + \frac{\varepsilon_{yy}}{\varepsilon_{zz}} k_y^2 + k_z^2 = \left(\frac{\omega}{c}\right)^2, \tag{23a}$$

$$\frac{1}{\varepsilon_{yy}\mu_{zz}} k_x^2 + \frac{1}{\varepsilon_{xx}\mu_{zz}} k_y^2 + k_z^2 = \left(\frac{\omega}{c}\right)^2. \tag{23b}$$

Here $\mathbf{k} = k_x \hat{\mathbf{x}} + k_y \hat{\mathbf{y}} + k_z \hat{\mathbf{z}}$ is the wave vector associated with a propagation factor of the type $e^{i\mathbf{k}\cdot\mathbf{r}}$. The polarization decomposition is a consequence of the fact that $\bar{\mu}^{-1} \cdot \bar{\varepsilon}$ has the same structure as in a uniaxial material, specifically it has two degenerate eigenvalues [63]. Due to this reason, we refer to this class of materials as a "pseudo-uniaxial" media.

From Eq. (23) it is readily seen that for propagation along $z$, the dispersion reduces to $k_z = \frac{\omega}{c}$ for both polarizations, analogous to the case of Minkowskian isotropic media (MIM) with $n = 1$ [55, 62]. However, for oblique propagation directions, the refractive index seen by the wave can be rather different from $n = 1$. To illustrate this, we depict in Fig. 6b the isofrequency surfaces of a material with $\varepsilon_{xx} = 1/\mu_{yy} = 8$, $\varepsilon_{yy} = 1/\mu_{xx} = 16$ and $\varepsilon_{zz} = \mu_{zz} = 1$. They consist of two different ellipsoids that touch along the $z$-axis at the points $k_z = \pm\frac{\omega}{c}$, consistent with Eq. (23).

For comparison, we show in Fig. 6a the dispersion of a MIM with the material parameters $\varepsilon = 8, \mu = 1/8$. As seen, in this case the isofrequency surfaces of the MIM are spherical and degenerate. They coincide with the vacuum dispersion ($k = \omega/c$). We note in passing that the MIM concept can be extended to other isorefractive materials with $n = const. > 1$ through the use of generalized Lorentz transformations [55]. For simplicity, in this article we only discuss systems invariant under a standard Lorentz boost.

It is simple to explain why the refraction index along the direction of the boost must satisfy $n = 1$. In fact, a non-dispersive medium response can remain invariant under a boost directed along $z$, only if $k_z/\omega = k'_z/\omega'$ for waves propagating along $z$. Using a relativistic Doppler transformation [28], we find that

$$\frac{\omega'}{k'_z} = \frac{\gamma(\omega - k_z v)}{\gamma\left(k_z - v\frac{\omega}{c^2}\right)} = \frac{c}{n}\left(\frac{1 - \frac{v}{c}n}{1 - \frac{v}{c}\frac{1}{n}}\right), \tag{24}$$

with $n/c = k_z/\omega$. The above equation is compatible with $k_z/\omega = k'_z/\omega'$ only when $n = 1$.

In the final example, we consider media such that $\varepsilon_{yy}\varepsilon_{xx} = \varepsilon_{zz}/\mu_{zz}$. From Eq. (23), one can see that the two dispersion equations become identical, analogous to the isotropic case, although the isofrequency contours are in general ellipsoids. We refer to such systems as "pseudo-isotropic" ($\overline{\mu}^{-1} \cdot \overline{\varepsilon}$ is a scalar). The dispersion of a pseudo-isotropic medium with the parameters $\varepsilon_{xx} = 1/\mu_{yy} = 8$, $\varepsilon_{yy} = 1/\mu_{xx} = 16$ and $\varepsilon_{zz} = \varepsilon_{xx}, \mu_{zz} = \mu_{xx}$ is represented in Fig. 6c. In this example, the isofrequency contours in the $xoz$ plane are circles [see the inset of Fig. 6c shaded in blue color] because $\varepsilon_{xx}/\varepsilon_{zz} = 1$.

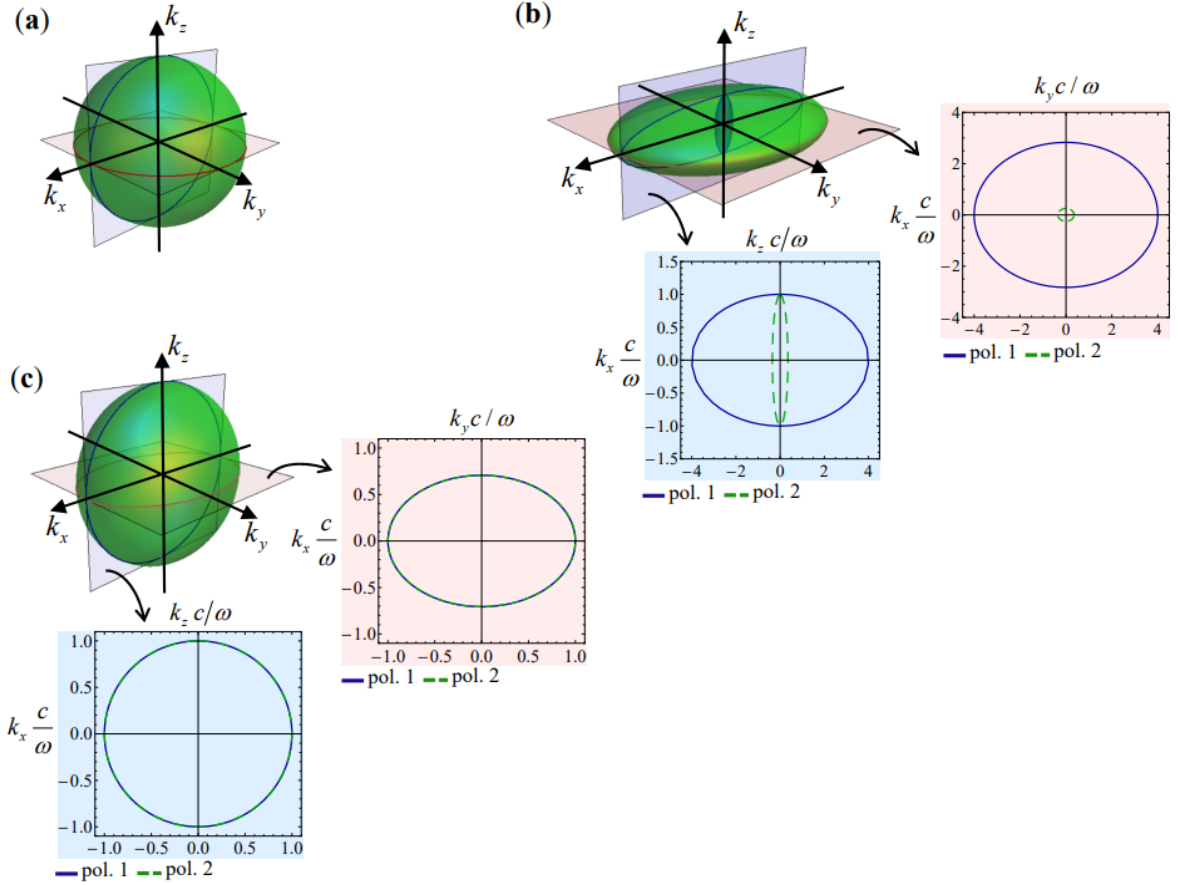

**Fig. 6** Isofrequency surfaces in *k*-space for **(a)** a Minkowskian isotropic material invariant under arbitrary Lorentz boosts. The material parameters are $\varepsilon = 8$ and $\mu = 1/8$, **(b)** a generalized (pseudo-uniaxial) Minkowskian medium invariant under Lorentz boosts along the *z*-direction. The material parameters are $\varepsilon_{xx} = 1/\mu_{yy} = 8$, $\varepsilon_{yy} = 1/\mu_{xx} = 16$ and $\varepsilon_{zz} = \mu_{zz} = 1$. **(c)** Similar to (b) but for a generalized (pseudo-isotropic) Minkowskian medium with the parameters $\varepsilon_{xx} = 1/\mu_{yy} = 8$, $\varepsilon_{yy} = 1/\mu_{xx} = 16$ and $\varepsilon_{zz} = \varepsilon_{xx}, \mu_{zz} = \mu_{xx}$. The insets depict the isofrequency contours in the *xoz* plane (shaded in blue) and in the *xoy* plane (shaded in pink).

### B. Band structure of generalized Minkowskian crystals

Next, we consider a spacetime modulated crystal formed by generalized Minkowskian materials invariant under boosts aligned with the *z*-direction. Due to the same reasons as in Ref. [55], such time-variant systems have an electromagnetic response indistinguishable from that of a moving photonic crystal. Due to the absence of magneto-electric coupling in the co-moving frame, the photonic band structure $\omega'$ vs. $\mathbf{k}'$ in this frame can be found using standard electromagnetic solvers. Then, the dispersion of the Bloch waves in the lab frame ($\omega$ vs. $\mathbf{k}$) can be found using the relativistic Doppler transformation [28, 55]:

$$k_x = k'_x, \quad k_y = k'_y, \quad k_z = \gamma\left(k'_z + v\frac{\omega'}{c^2}\right) \quad \text{and} \quad \omega = \gamma(\omega' + k'_z v). \quad (25)$$

Here, $\mathbf{k} = (k_x, k_y, k_z)$ and $\mathbf{k}' = (k'_x, k'_y, k'_z)$ are the wavevectors in the laboratory and co-moving frames, respectively.

To illustrate these ideas, first we consider a travelling wave crystal formed by MIM spheroid inclusions with permittivity and permeability given by $\varepsilon = 8$ and $\mu = 1/8$. The inclusions are subjected to a modulation velocity $v = 0.2c$ along the $z$-direction. In the co-moving frame, the inclusions have a spherical shape with radius $R = 0.35a$ and the same electromagnetic response as in the lab frame ($\varepsilon' = 8$, $\mu' = 1/8$). The exact co-moving frame band diagram of the 3D crystal (Fig. 7ai) was numerically computed using the electromagnetic simulator CST Microwave Studio [64]. For simplicity, we restrict our analysis to the case of Bloch modes that propagate along the $z$-direction, i.e., parallel to the modulation velocity. As expected, the co-moving frame dispersion exhibits spectral symmetry $\omega'(-k'_z) = \omega'(k'_z)$, due to the absence of a bianisotropic response in the co-moving frame. Figure 7aii shows the corresponding dispersion diagram in the lab frame obtained using the relativistic Doppler transformation [Eq. (25)]. Now, the spectral symmetry is broken $\omega(k_z) \neq \omega(-k_z)$, resulting in a synthetic Fresnel drag effect [20, 21, 55]. Specifically, the wave velocities in the laboratory frame $v^\pm = \omega/k_z^\pm$ can be determined from the wave velocities $v'^\pm = \omega'/k_z'^\pm$ in the co-moving frame using a relativistic addition of velocities [28, 55]:

$$v^\pm = \frac{v'^\pm + v}{1 + v'^\pm v/c^2}. \quad (26)$$

Note that in the co-moving frame $v'_+ = -v'_- \equiv v'$ due to the previously discussed spectral symmetry. The wave velocities coincide with the slopes of the photonic dispersion, represented with dashed lines in Figs. 7ai and 7aii. Due to the synthetic Fresnel drag effect, the wave velocities in the lab frame satisfy $|v^-| < v' < |v^+|$. This means that the waves that propagate in the direction of the crystal modulation travel faster than the waves that propagate in the opposite direction. Figure 7b shows a detailed study of the variation of $v^\pm$ with the modulation speed. As seen, for a sufficiently large $v$ the two velocities can have the same sign so that the propagation becomes effectively unidirectional [23].

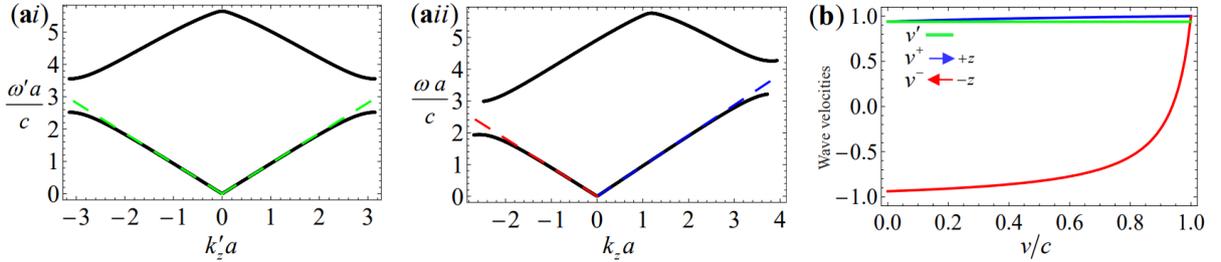

**Fig. 7** Exact dispersion diagram of a MIM 3D spacetime crystal [Fig. 1] for $\varepsilon = \varepsilon' = 8$, $\mu = \mu' = 1/8$ for a modulation speed $v = 0.2c$ calculated in the (a*i*) co-moving frame, using the electromagnetic simulator CST Microwave studio, and (a*ii*) laboratory frame, using a Doppler transformation [Eq. (25)]. (b) Long wavelength limit wave velocities in the laboratory frame $v^+, v^-$ as a function of the modulation speed $v/c$. The horizontal green line represents the wave velocity in the co-moving frame $|v'| = |v'^\pm|$.

### C. Validation of the Clausius-Mossotti homogenization

In order to validate our Clausius-Mossotti homogenization, we compare the refractive indices obtained from the slopes of the exact lab frame dispersion diagram $n^\pm = c|k_z^\pm|/\omega$ with the refractive indices calculated using the homogenization theory ($n_{\text{ef}}^\pm$). The latter are found by numerically the equation:

$$\det(\mathbf{N} - c\mathbf{M}_{\text{ef}}) = 0, \quad \mathbf{N} = \begin{pmatrix} \mathbf{0}_{3\times 3} & -\dfrac{c\mathbf{k}}{\omega} \times \mathbf{1}_{3\times 3} \\ +\dfrac{c\mathbf{k}}{\omega} \times \mathbf{1}_{3\times 3} & \mathbf{0}_{3\times 3} \end{pmatrix}, \qquad (27)$$

with $\mathbf{k} = k_z\hat{\mathbf{z}}$ and $\mathbf{M}_{ef}$ defined as in Eq. (12).

In the first example, we consider a Minkowskian crystal with the same parameters as in Fig. 7aii, but with varying inclusion radius. Figure 8a shows the lab frame refractive indices computed with the help of CST Microwave Studio simulations and the Doppler transformation [Eq. (26)], and the corresponding homogenization results for propagation along +z and –z directions. There is an excellent agreement between the two calculation methods. Note that in this example the two waves that propagate along +z are degenerate. Similarly, the two waves that propagate along –z are also degenerate.

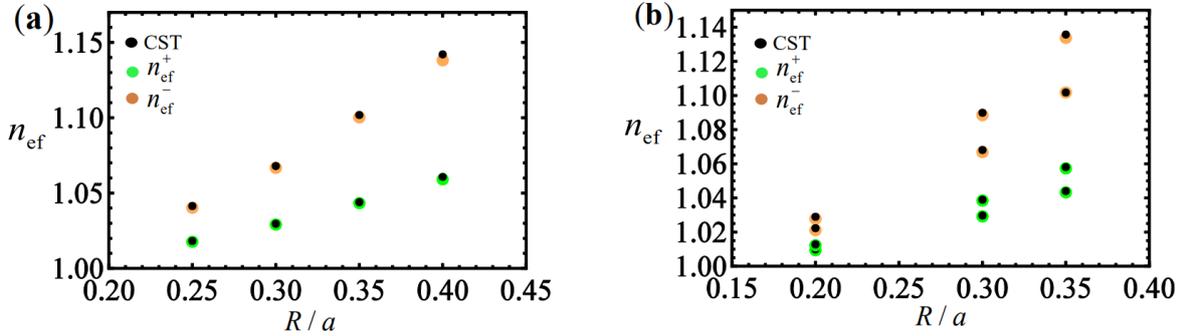

**Fig. 8** Exact lab-frame effective refractive indices calculated using CST Microwave Studio [64] (black points) and refractive indices predicted by the homogenization theory (green and dark orange points) [Eq. (27)] for different values of the radius $R$. The points associated with propagation along +z are represented with green circles, whereas the points corresponding to propagation along –z are represented with orange circles. The modulation speed is $v = 0.2c$. **(a)** Minkowskian spacetime crystal formed by MIM inclusions, with $\varepsilon = 8$ and $\mu = 1/8$. **(b)** Generalized Minkowskian spacetime crystal formed by anisotropic inclusions, with $\varepsilon_{xx} = 1/\mu_{yy} = 8$, $\varepsilon_{yy} = 1/\mu_{xx} = 16$ and $\varepsilon_{zz} = \mu_{zz} = 1$, for the two polarizations.

In the second example, we characterize the effective response of a pseudo-uniaxial generalized Minkowskian crystal. The material parameters of the inclusions are the same as in Fig. 6b. As seen in Fig. 8b, the homogenization result matches well the exact refractive indices. In this example, the effective medium has two different refractive indices per propagation direction due to the anisotropy of the inclusions material. Thus, the synthetic Fresnel drag is sensitive to the wave polarization.

Note that in the first example, the effective response is characterized by a moving medium bianisotropic coupling. On the other hand, in the second example the bianisotropic response is more complex ($\xi_{\text{ef},11} = \xi_{\text{ef},22} = 0$ and $\xi_{\text{ef},12} \neq -\xi_{\text{ef},21}$) due to the anisotropy of the inclusions.

## IV. Conclusion

In this work, we have developed a comprehensive analytical formalism to characterize the effective electromagnetic response of 3D spacetime crystals formed by spherical scatterers under a travelling-wave modulation. By extending the Clausius-Mossotti formula to time-varying platforms, we have shown that these spacetime crystals can be engineered to exhibit a wide range of nonreciprocal bianisotropic couplings in the long-wavelength limit.

Importantly, we demonstrated the possibility of achieving a purely isotropic Tellegen response in spacetime crystals formed by interlaced sub-lattices of scatterers subjected to different modulation velocities, introducing a novel mechanism to engineer isotropic nonreciprocal responses. In addition, we introduced a new class of materials that display invariance under Lorentz boosts along a fixed spatial direction. These pseudo-uniaxial spacetime crystals have an electromagnetic response indistinguishable from that of a moving photonic crystal, which simplifies the analysis of wave propagation and broadens the understanding of spacetime crystal behavior under relativistic transformations. Finally, our theoretical predictions were supported by numerical simulations, demonstrating that the homogenization theory accurately describes the wave dispersion in 3D Minkowskian spacetime crystals. Our analysis opens new pathways for engineering advanced nonreciprocal materials and devices with tailored electromagnetic responses.

**Acknowledgements:** This work is supported in part by the IET under the A F Harvey Engineering Research Prize, by the Simons Foundation, by Fundação para a Ciência e a Tecnologia under project 2022/06797/PTDC, and by Instituto de Telecomunicações under project UIDB/50008/2020.

## Appendix A: Polarizability of a bianisotropic particle

The polarizability of the spherical bianisotropic particle in the co-moving frame can be found with a quasi-static approximation. In the quasi-static limit, the electric and magnetic fields are effectively decoupled and can be written in terms of electric and magnetic potentials as follows: $\mathbf{E} = -\nabla \phi_e$ and $\mathbf{H} = -\nabla \phi_m$. It is implicit that the fields are evaluated in the co-moving frame coordinates (for notational simplicity we omit the primes in the fields in this Appendix).

In the quasi-static limit, it is necessary that $\nabla \cdot \mathbf{D} = 0 = \nabla \cdot \mathbf{B}$. In the air region, it is evident that these two equations reduce to the Laplace equation for the potentials: $\nabla^2 \phi_e = 0 = \nabla^2 \phi_m$. Furthermore, we look for a solution such that $\mathbf{E} = \mathbf{E}_1 = const.$ and $\mathbf{H} = \mathbf{H}_1 = const.$ inside the bianisotropic particle. The equations $\nabla \cdot \mathbf{D} = 0 = \nabla \cdot \mathbf{B}$ are then automatically satisfied inside a uniform particle described by a material matrix $\mathbf{M}$ independent of the spatial coordinates.

Based on these considerations, we consider the following ansatz for the electric and magnetic potentials (see also Refs. [65, 66]):

$$\phi_e = \begin{cases} -\mathbf{E}^{inc} \cdot \mathbf{r} + \dfrac{\mathbf{p} \cdot \hat{\mathbf{r}}}{4\pi\varepsilon_0 r^2}, & r > R \\ -\mathbf{E}_1 \cdot \mathbf{r}, & r < R \end{cases} \quad \text{(A1a)}$$

$$\phi_m = \begin{cases} -\mathbf{H}^{inc} \cdot \mathbf{r} + \dfrac{\mathbf{m} \cdot \hat{\mathbf{r}}}{4\pi r^2}, & r > R \\ -\mathbf{H}_1 \cdot \mathbf{r}, & r < R \end{cases} \quad \text{(A1b)}$$

Here, $\mathbf{E}^{inc}, \mathbf{H}^{inc}$ represent the incident fields created by some external excitation and $\mathbf{p}, \mathbf{m}$ are the (unknown) electric and magnetic dipoles induced on the bianisotropic particle. In order to find, $\mathbf{p}, \mathbf{m}$ and $\mathbf{E}_1, \mathbf{H}_1$ we enforce boundary conditions at the interface of the spherical particle.

The continuity of the potentials $\phi_e, \phi_m$ leads to:

$$\begin{pmatrix} \mathbf{E}_1 \\ \mathbf{H}_1 \end{pmatrix} = \begin{pmatrix} \mathbf{E}^{inc} \\ \mathbf{H}^{inc} \end{pmatrix} - \begin{pmatrix} \dfrac{\mathbf{p}}{4\pi\varepsilon_0 R^3} \\ \dfrac{\mathbf{m}}{4\pi R^3} \end{pmatrix}. \tag{A2}$$

On the other hand, the continuity of the normal components of the electric displacement ($\mathbf{D}$) and magnetic induction ($\mathbf{B}$) vectors requires that:

$$\mathbf{M}_0 \cdot \begin{pmatrix} \mathbf{E}^{inc} \\ \mathbf{H}^{inc} \end{pmatrix} + \mathbf{M}_0 \cdot \begin{pmatrix} \dfrac{2\mathbf{p}}{4\pi\varepsilon_0 R^3} \\ \dfrac{2\mathbf{m}}{4\pi R^3} \end{pmatrix} = \mathbf{M}' \cdot \begin{pmatrix} \mathbf{E}_1 \\ \mathbf{H}_1 \end{pmatrix}. \tag{A3}$$

Here, $\mathbf{M}_0$ stands for the material matrix of the vacuum and $\mathbf{M}'$ is the material matrix of the bianisotropic particle, defined as in the main text. Combining Eqs. (A2)-(A3), one gets:

$$\mathbf{M}_0 \cdot \begin{pmatrix} \mathbf{E}^{inc} \\ \mathbf{H}^{inc} \end{pmatrix} + \frac{1}{2\pi R^3} \mathbf{M}_0 \cdot \begin{pmatrix} \mathbf{p}/\varepsilon_0 \\ \mathbf{m} \end{pmatrix} = \mathbf{M}' \cdot \begin{pmatrix} \mathbf{E}^{inc} \\ \mathbf{H}^{inc} \end{pmatrix} - \frac{1}{4\pi R^3} \mathbf{M}' \cdot \begin{pmatrix} \mathbf{p}/\varepsilon_0 \\ \mathbf{m} \end{pmatrix}. \tag{A4}$$

Solving the above equation with respect to the dipole moments, one finally finds that:

$$\begin{pmatrix} \mathbf{p}/\varepsilon_0 \\ \mathbf{m} \end{pmatrix} = \boldsymbol{\alpha} \cdot \begin{pmatrix} \mathbf{E}^{inc} \\ \mathbf{H}^{inc} \end{pmatrix}, \quad \text{with} \quad \boldsymbol{\alpha} = \alpha_0 \left( \mathbf{M}' + 2\mathbf{M}_0 \right)^{-1} \cdot \left( \mathbf{M}' - \mathbf{M}_0 \right) \tag{A5}$$

and $\alpha_0 = 4\pi R^3$.

## Appendix B: Tensor components for a moving-Tellegen effective response

Here, we present explicit formulas for the elements of the tensors in Eqs. (16). It is assumed that the principal axes of the permeability are aligned with the $x$ and $y$ directions, $\hat{\mathbf{u}}_1 \equiv \hat{\mathbf{x}}, \hat{\mathbf{u}}_2 \equiv \hat{\mathbf{y}}$, and that $\hat{\mathbf{u}}_3 \equiv \hat{\mathbf{z}}$.

Using the relations, $\hat{\mathbf{e}}_1 = \cos(\theta)\hat{\mathbf{u}}_1 + \sin(\theta)\hat{\mathbf{u}}_2$ and $\hat{\mathbf{e}}_2 = -\sin(\theta)\hat{\mathbf{u}}_1 + \cos(\theta)\hat{\mathbf{u}}_2$ [Fig. 3a], it is straightforward to show that the effective permittivity is given by:

$$\bar{\varepsilon}_{\mathrm{ef}} = \begin{pmatrix} 1+\Delta_\varepsilon+\cos(2\theta)\delta_\varepsilon f_V & \sin(2\theta)\delta_\varepsilon f_V & 0 \\ \sin(2\theta)\delta_\varepsilon f_V & 1+\Delta_\varepsilon-\cos(2\theta)\delta_\varepsilon f_V & 0 \\ 0 & 0 & 1 \end{pmatrix}, \tag{B1}$$

whereas the effective permeability satisfies:

$$\bar{\mu}_{\mathrm{ef}} = \begin{pmatrix} 1+\Delta_\mu+\delta_\mu f_V & 0 & 0 \\ 0 & 1+\Delta_\mu-\delta_\mu f_V & 0 \\ 0 & 0 & 1 \end{pmatrix}. \tag{B2}$$

Finally, the magneto-electric tensor is given by:

$$\bar{\xi}_{\mathrm{ef}} = \begin{pmatrix} \xi\sin 2\theta & \xi\cos 2\theta & 0 \\ -\xi\cos 2\theta & \xi\sin 2\theta & 0 \\ 0 & 0 & 0 \end{pmatrix}. \tag{B3}$$

## Appendix C: Minkowskian systems

Under a Lorentz transformation the material matrix is transformed as $\mathbf{M} \to \mathbf{M}'$, with $\mathbf{M}'$ given by Eq. (4) of the main text. In this Appendix, we characterize the "fixed points" of this transformation for a given boost direction.

For simplicity, we solve the problem perturbatively assuming that $\mathbf{M} = \mathbf{M}_0 + \delta\mathbf{M}$ with $\mathbf{M}_0$ the material matrix of the vacuum and $\delta\mathbf{M}$ a small perturbation. By substituting $\mathbf{M} = \mathbf{M}_0 + \delta\mathbf{M}$ in Eq. (4), we find that the transformed matrix, to leading order in the small parameter $\delta\mathbf{M}$ is given by:

$$\begin{aligned}\mathbf{M}' &\approx \mathbf{A}\cdot\delta\mathbf{M}\cdot[\mathbf{A}+\mathbf{V}\cdot\mathbf{M}_0]^{-1}+\left[\frac{1}{c^2}\mathbf{V}+\mathbf{A}\cdot\mathbf{M}_0\right]\cdot[\mathbf{A}+\mathbf{V}\cdot\mathbf{M}_0+\mathbf{V}\cdot\delta\mathbf{M}]^{-1}\\ &\approx \mathbf{A}\cdot\delta\mathbf{M}\cdot[\mathbf{A}+\mathbf{V}\cdot\mathbf{M}_0]^{-1}+\left[\frac{1}{c^2}\mathbf{V}+\mathbf{A}\cdot\mathbf{M}_0\right]\cdot\left[\mathbf{1}-(\mathbf{A}+\mathbf{V}\cdot\mathbf{M}_0)^{-1}\cdot\mathbf{V}\cdot\delta\mathbf{M}\right][\mathbf{A}+\mathbf{V}\cdot\mathbf{M}_0]^{-1}\\ &\approx \mathbf{M}_0+\mathbf{A}\cdot\delta\mathbf{M}\cdot[\mathbf{A}+\mathbf{V}\cdot\mathbf{M}_0]^{-1}-\mathbf{M}_0\cdot\mathbf{V}\cdot\delta\mathbf{M}\cdot[\mathbf{A}+\mathbf{V}\cdot\mathbf{M}_0]^{-1}.\end{aligned}$$
(C1)

In the second identity, we used $[\mathbf{1}+\mathbf{B}]^{-1}\approx\mathbf{1}-\mathbf{B}$ for small $\mathbf{B}$, and in the third identity we took into account that vacuum is invariant under any Lorentz boost, $\mathbf{M}_0=\left[\frac{1}{c^2}\mathbf{V}+\mathbf{A}\cdot\mathbf{M}_0\right]\cdot[\mathbf{A}+\mathbf{V}\cdot\mathbf{M}_0]^{-1}$. Equation (C1) shows that $\mathbf{M}'\approx\mathbf{M}_0+\delta\mathbf{M}'$ with

$$\delta\mathbf{M}'\approx[\mathbf{A}-\mathbf{M}_0\cdot\mathbf{V}]\cdot\delta\mathbf{M}\cdot[\mathbf{A}+\mathbf{V}\cdot\mathbf{M}_0]^{-1}. \quad (C2)$$

It can be checked that $[\mathbf{A}-\mathbf{V}\cdot\mathbf{M}_0]=[\mathbf{A}+\mathbf{V}\cdot\mathbf{M}_0]^{-1}$. Thus, a perturbation with respect to the vacuum $\mathbf{M}=\mathbf{M}_0+\delta\mathbf{M}$ in certain frame leads to a perturbation in another frame $\mathbf{M}'\approx\mathbf{M}_0+\delta\mathbf{M}'$, such that

$$\delta\mathbf{M}'\approx[\mathbf{A}-\mathbf{M}_0\cdot\mathbf{V}]\cdot\delta\mathbf{M}\cdot[\mathbf{A}-\mathbf{V}\cdot\mathbf{M}_0]. \quad (C3)$$

Retaining only terms that are linear in the velocity, we obtain that

$$\delta\mathbf{M}'\approx\delta\mathbf{M}-\mathbf{M}_0\cdot\tilde{\mathbf{V}}\cdot\delta\mathbf{M}-\delta\mathbf{M}\cdot\tilde{\mathbf{V}}\cdot\mathbf{M}_0, \quad (C4)$$

with $\tilde{\mathbf{V}}=\begin{pmatrix}0 & \mathbf{v}\times\mathbf{1}\\ -\mathbf{v}\times\mathbf{1} & 0\end{pmatrix}$. So, the perturbed vacuum can be invariant under the Lorentz boost only if

$$\mathbf{M}_0\cdot\tilde{\mathbf{V}}\cdot\delta\mathbf{M}+\delta\mathbf{M}\cdot\tilde{\mathbf{V}}\cdot\mathbf{M}_0=0. \quad (C5)$$

Here, we suppose that $\delta\mathbf{M}$ models a non-bianisotropic material, so that:

$$\delta\mathbf{M}=\begin{pmatrix}\varepsilon_0\delta\boldsymbol{\varepsilon} & \mathbf{0}_{3\times 3}\\ \mathbf{0}_{3\times 3} & \mu_0\delta\boldsymbol{\mu}\end{pmatrix}. \quad (C6)$$

In this case, Eq. (C5) reduces to:

$$\mathbf{v}\times\delta\boldsymbol{\mu}+\delta\boldsymbol{\varepsilon}\cdot\left[\mathbf{v}\times\mathbf{1}_{3\times 3}\right]=0, \qquad \delta\boldsymbol{\mu}\cdot\left[\mathbf{v}\times\mathbf{1}_{3\times 3}\right]+\mathbf{v}\times\delta\boldsymbol{\varepsilon}=0. \tag{C7}$$

The above equations can be regarded as a linear system in $\delta\boldsymbol{\varepsilon},\delta\boldsymbol{\mu}$. Imposing that these equations are satisfied for boosts directed along the $z$-direction, it is found that to leading order:

$$\delta\boldsymbol{\varepsilon}=\begin{pmatrix}\delta\varepsilon_{xx} & \delta\varepsilon_{xy} & 0\\ \delta\varepsilon_{yx} & \delta\varepsilon_{yy} & 0\\ 0 & 0 & \delta\varepsilon_{zz}\end{pmatrix}, \quad \delta\boldsymbol{\mu}=\begin{pmatrix}-\delta\varepsilon_{yy} & \delta\varepsilon_{yx} & 0\\ \delta\varepsilon_{xy} & -\delta\varepsilon_{xx} & 0\\ 0 & 0 & \delta\mu_{zz}\end{pmatrix}. \tag{C8}$$

The above conditions guarantee Minkowskian invariance for small $\delta\boldsymbol{\varepsilon},\delta\boldsymbol{\mu}$ and $v/c \ll 1$ with the boost velocity directed along $z$.

The equations can be generalized to arbitrary values of $v/c$ and for arbitrarily large values of $\delta\boldsymbol{\varepsilon},\delta\boldsymbol{\mu}$ as shown in Eq. (21). The condition $\mathbf{M}'=\mathbf{M}$ can be verified by direction substitution of Eq. (21) into Eq. (4) and holds true even when $\delta\boldsymbol{\varepsilon},\delta\boldsymbol{\mu}$ are not Hermitian. The details are omitted by conciseness. Note that for small $\delta\boldsymbol{\varepsilon},\delta\boldsymbol{\mu}$ Eq. (21) is consistent with the perturbations shown in Eq. (C8).